\newcommand{\PLA}[3]{Phys.\ Lett.\ A\ {\bf #1},\ #2 (#3)}

\newcommand{\PRL}[3]{Phys.\ Rev.\ Lett.\ {\bf #1},\ #2 (#3)}

\newcommand{\NAT}[3]{Nature\ {\bf #1},\ #2 (#3)}

\newcommand{\PRA}[3]{Phys.\ Rev.\ A\ {\bf #1},\ #2 (#3)}

\newcommand{\JETP}[3]{JETP.\ Lett.\ {\bf #1},\ #2 (#3)}

\newcommand{\CTP}[3]{Commun.\ Theor.\ Phys.\ {\bf #1},\ #2 (#3)}
\newcommand{\JKP}[3]{J.\ Kor.\ Phys.\ Soc.\  {\bf #1},\ #2 (#3)}
\newcommand{\Chtp}[3]{Chin.\ Phys.\ B. \  {\bf #1},\ #2 (#3)}

\documentclass[aps,showkeys,showpacs,amsmath,amssymb]{revtex4}
\usepackage{dcolumn}
\usepackage{longtable}
\begin{document}
\title{Quantum information splitting using multi-partite cluster states}
\author{Sreraman Muralidharan}
\email{sreraman_m@yahoo.co.in} \affiliation{Loyola College,
Nungambakkam, Chennai - 600 034, India}
\author{Prasanta K. Panigrahi}
\email{prasanta@prl.res.in}
\affiliation{Indian Institute of Science Education and Research (IISER) Kolkata, Salt Lake, Kolkata - 700106, India}
\affiliation{Physical Research
Laboratory, Navrangpura, Ahmedabad - 380 009, India}
\begin{abstract}
We provide various schemes for the splitting up of Quantum information into parts using the 
four and five partite cluster states. Explicit protocols for the Quantum
information splitting (QIS) of single and two qubit states are illustrated. It is found that the four partite
cluster state can be used for the QIS of an entangled state and the five partite cluster
state can be used for QIS of an arbitrary two qubit state. The schemes considered here are also
secure against certain eavesdropping attacks. 
\end{abstract}


\maketitle
\section{Introduction}
Quantum entanglement, a feature not available in classical physics, has led to many advances 
in communication theory and cryptography. It has found many practical
applications in teleportation, dense coding and secret sharing. Quantum teleportation is a technique for transfer of information
between parties, using a distributed entangled state and a
classical communication channel \cite{Bennett}.  
It also serves as an elementary operation for a number of quantum communicational protocols
and has been experimentally demonstrated for arbitrary single and
two qubit systems using pairs of Bell states \cite{tele1, Bsm}.  With better understanding of multipartite entanglement, several authors have devised protocols for teleportation using multiparticle entangled channels \cite{ Chuang28, Gorbo, Shi, Deng28, Rigolin28, Sreramanpr}. \\

Quantum secret sharing (QSS) is the generalization of classical secret sharing schemes to the quantum scenario \cite{Gotit}. 
In QSS is that the owner who possesses and wishes to transmit the secret information
splits it among various parties such that the original
information can only be reconstructed by a specific subset of the parties. QSS plays
a key role in transmitting and protecting both classical as well as quantum information and hence it can be
further divided into two branches, namely secret sharing of classical information and quantum information.\\

Hillery \it{et al.} \normalfont \cite{Hillery}, described
the first scheme for secret sharing of quantum information by extending quantum key sharing to quantum information splitting
by the use of teleportation. In their scheme, the three parties, namely Alice, Bob and Charlie share an entangled three qubit
GHZ state :
\begin{equation}
|GHZ\rangle_{ABC} = \frac{1}{\sqrt{2}} (|000\rangle \pm |111\rangle)_{ABC},
\end{equation}
each possessing one qubit. Alice has an unknown single qubit information, $|\psi_a\rangle = \alpha|0\rangle+\beta|1\rangle$, which 
she wants Bob and Charlie to share. Initially, Alice combines $|\psi_a\rangle$ with
her qubit in the entangled state and performs a Bell measurement on her pair of qubits and conveys the outcome of her
measurement to Charlie via two classical bits, after which the Bob-Charlie system collapses
into an entangled state given by
\begin{equation}
U_x \otimes I (\alpha|00\rangle+\beta|11\rangle) _{BC},
\end{equation} 
where $U_x \in ({I, \sigma_x, i\sigma_y, \sigma_z})$. Hence, the unknown single qubit information is locked between Bob and Charlie
in such a way that neither of them can obtain the unknown qubit completely, by locally operating on their own qubits. 
Now, Bob can perform a single qubit measurement in the basis $\frac{1}{\sqrt{2}}(|0\rangle\pm|1\rangle)$
on his qubit and convey the outcome of his measurement to Charlie via one classical bit. Having, known
the outcomes of both their measurements, Charlie can obtain $|\psi_a\rangle$ by performing
a suitable unitary operation. This technique of splitting and sharing of
quantum information among two or more parties such that none of them can retrieve
the information fully by operating on their own qubits, is usually referred to as "Quantum information splitting (QIS)". 
The QIS of $|\psi_a\rangle$ involving the tri-partite $|GHZ\rangle_3$ as an entangled channel along with many other probabilistic schemes has been extensively studied and discussed in \cite{Band}. The in-principle feasibility of experimental realization of QIS of $|\psi_a\rangle$ has been demonstrated through a pseudo GHZ state \cite{Tittel28}. Later, an experimental scheme using single photon sources for splitting up of $|\psi_a\rangle$ was demonstrated \cite{Christ}. Recently, attention has turned towards the usage of different types of multipartite entangled channels for QIS. 
	Four party secret sharing has been experimentally shown using, a special four photon
polarization entangled state \cite{Gaer}.
QIS of $|\psi_a\rangle$ has also been carried out using an asymmetric W state \cite{Zheng} given by,
\begin{equation}
|W_a\rangle = (\frac{1}{2} |001\rangle + \frac{1}{2} |010\rangle + \frac{1}{\sqrt{2}} |001\rangle)_{ABC}
\end{equation}
and experimentally realized in ion-trap systems. 
Later, schemes have been devised, which used cavity QED to split $|\psi_a\rangle$ via $|W_a\rangle$ \cite{Wqed, Wqed2}. 
Many other QIS schemes have been proposed, in which the particle carrying
the information that is to be teleported, needs to be initially entangled with other particles \cite{Karl, Cleve}.
Recently, a generalization of QIS has been made from qubits to qudits \cite{quditqis}. With the increase in the number of qubits, the scenario turns complicated, owing to the lack of proper understanding of multi-particle entanglement.

The QIS of an arbitrary two qubit state given by
\begin{equation}
|\psi_b\rangle=\alpha|00\rangle+\mu|10\rangle+\gamma|01\rangle+\beta|11\rangle,
\end{equation} 
was initially carried out using four Bell pairs as an entangled resource \cite{Deng}. 
Later, we proposed a robust scheme \cite{Sreraman},
 involving lesser number of particles for the 
 splitting up of $|\psi_a\rangle$ and $|\psi_b\rangle$. In this paper,
we shall present alternate schemes for splitting up of arbitrary single and two 
qubit states using four and five partite cluster states and discuss its advantages over all the
schemes known before. Upto six partite cluster states have been experimentally realized in laboratory
conditions.  These states were first introduced in Ref. \cite{Robert} for linear optics
one way quantum computation.  Like the previous ones, our schemes are also secure against certain eavesdropping attacks. \\
  
In general, an $N$ qubit cluster state is given by \cite{Robert} :
\begin{equation}
|C_N\rangle = \frac{1}{2^{N/2}} \otimes_{a=1}^{N} (|0\rangle_a \sigma_z ^{a+1} + |1\rangle_a), 
\end{equation}
with $\sigma_z^{N+1} = 1$. They show a strong violation of local reality and are shown to be
robust against decoherence \cite{Walther, Hein}.  Much work has been done
in trying to characterize the entanglement exhibited by these states, owing to
their promising usefulness in Quantum information theory \cite{Bai, MEMS}.
These states have been identified as Task-oriented Maximally Entangled States
\cite{Pankaj}.  In the case of two and three partite scenarios, the cluster state is same as the Bell and the GHZ states respectively, under LOCC and for higher qubit systems, the state is not locally equivalent to the GHZ states and exhibit remarkably different entanglement properties \cite{Robert}. Recently, upto six particle cluster states have been experimentally
created \cite{Lu}. This motivates us to investigate the usefulness of these states for QIS of
single and two qubit states. \\

The schemes proposed here does not require the 
initial qubit to be entangled with other qubits, as in Ref. \cite{Karl}.
Hence it is as useful as the QIS schemes involving the $|GHZ_3\rangle$ and $|W_a\rangle$ states for splitting
up of $|\psi_a\rangle$. Moreover, any $N$ -partite GHZ and the asymmetric W states \cite{Zheng} cannot be used for the QIS
of an arbitrary two qubit state, while we show here that higher dimensional cluster states ($N>4$) can be used for this purpose. 
Previous scheme for the QIS of an arbitrary two qubit state required four Bell pairs, but the same is achieved here using
a five partite entangled state. Unlike the Brown state, cluster states have been realized in laboratory conditions. Hence, our schemes using a five qubit cluster state is more advantageous than the known schemes \cite{Deng, Sreraman} for the QIS of an arbitrary two qubit state. Like the previous ones, our schemes are also secure against certain types of eavesdropping attacks.   \\

This paper is organized as follows: In the following section, we devise protocols for 
splitting up of an unknown single and entangled two qubit states using $|C_4\rangle$ as an entangled channel. 
Subsequently in section III, we devise schemes for splitting of an unknown single and two qubit states
using $|C_5\rangle$ as an entangled channel. The conclusion follows in section IV, where we summarise
our scheme and point out the direction for future work. 

\section{$|C_4\rangle$ for QIS}
The four partite cluster state is given by,
\begin{eqnarray}
|C_4\rangle = \frac{1}{2}(|0000\rangle + |0110\rangle + |1001\rangle - |1111\rangle)_{1234}.
\end{eqnarray}
$|C_4\rangle$ has been proven to be of immense use in one way Quantum computation \cite{Robert, Hans} and also for Quantum error correction \cite{Sch}.  One way quantum computation has been experimentally demonstrated \cite{Walther}. There has been 
extensive investigation regarding the usefulness of these states for teleportation and dense coding \cite{Pankaj, 4pati}.
It has been shown that, this state could be used for perfect teleportation
of an arbitrary two qubit state and that its superdense coding capacity  
reaches the Holevo bound i.e., one can send four classical bits by sending only two quantum bits using
two ebits of entanglement \cite{Pankaj}. We shall now demonstrate the usefulness
of this state for QIS of single and two qubit systems.
\subsection{Single qubit state} 
We let Alice possess qubit 1, Bob possess qubits 2, 3 and Charlie 4. Alice combines $|\psi_a\rangle$ with her qubit in
the entangled state and performs a Bell measurement. The outcome of the measurement performed by Alice and the entangled state obtained by Bob and Charlie are shown in table \ref{tab2}.
\begin{table}[h]
\caption{\label{tab2} The outcome of the measurement performed by Alice and the state obtained by Bob and Charlie}
\begin{tabular}{|c|c|}
\hline {\bf Outcome of the Measurement } & {\bf State obtained }\\
$\frac{1}{\sqrt{2}}(|00\rangle+|11\rangle)_{a1}$&$(\alpha(|000\rangle+|110\rangle)+\beta(|001\rangle-|111\rangle))_{234}$\\
$\frac{1}{\sqrt{2}}(|00\rangle-|11\rangle)_{a1}$&$(\alpha(|000\rangle+|110\rangle)-\beta(|001\rangle-|111\rangle))_{234}$\\
$\frac{1}{\sqrt{2}}(|01\rangle+|10\rangle)_{a1}$&$(\alpha(|001\rangle-|111\rangle)+\beta(|000\rangle+|110\rangle))_{234}$\\
$\frac{1}{\sqrt{2}}(|01\rangle-|10\rangle)_{a1}$&$(\alpha(|001\rangle-|111\rangle)-\beta(|000\rangle+|110\rangle))_{234}$\\
\hline
\end{tabular}
\end{table}

Instead of a Bell measurement, Alice can also perform two single particle measurements in the basis $\frac{1}{\sqrt{2}}(|0\rangle\pm|1\rangle)$. 
Alice can communicate the results of her measurement to Charlie using two cbits of information. Bob then performs a measurement in the basis $({|00\rangle_{23}, |11\rangle}_{23})$ and communicates the outcome of his results to Charlie via one cbit of information. 
Having known the outcomes of both their measurements, Charlie can obtain $|\psi_a\rangle$ by applying a suitable unitary operator on his qubit. 
For instance, had the Bob-Charlie system evolved into the first state shown in table 2 and if the outcome of Bob's measurement is $|00\rangle_{23}$, then Charlie's state collapses to $(\alpha|0\rangle+\beta|1\rangle)_4$; Instead if the outcome of Bob's measurement is $|11\rangle_{23}$, then
Charlie's state collapses to $(\alpha|0\rangle-\beta|1\rangle)_4$. Charlie can obtain $|\psi_a\rangle$ by applying 
a suitable unitary operator on his qubit.\\

Now, let us discuss the security 
of this protocol against an eavesdropper (say Eve). We assume that, Eve has managed to entangle an ancillia to 
a qubit possessed by Bob in the four qubit cluster state, so that she can measure the ancillia to gain
information about the unknown qubit state. Suppose, all the three participants are unaware of this attack
of Eve, then after Alice performs a Bell measurement, the combined state of Bob, Charlie and Eve collapses into
a four-partite entangled state. However, after Bob-performs the two particle measurement, the Charlie-Eve system collapses 
into a product state, leaving Eve with no information about the unknown qubit. To see this scenario more explicitly, let us assume that Eve has managed
to entangle an ancillia $|0\rangle$ to the second qubit of the entangled channel. 
If Alice performs a measurement in the first basis, then the combined state of Bob, Charlie and Eve would be,
\begin{equation}
\alpha(|0000\rangle+|1101\rangle)+\beta(|0010\rangle-|1111\rangle) _ {BCE}.
\end{equation}
Now if Bob performs a measurement in the basis $|00\rangle$, then the Charlie-Eve system collapses to, $(\alpha|0\rangle+\beta|1\rangle)_{C} |0\rangle_{E}$. Hence, Eve's state is unaltered leaving no chance for her to gain any information about the unknown qubit state. This
is due to the fact that, entanglement is monogamous \cite{mono}.    
\subsection{Two qubit state}
The $|C_4\rangle$ state can be used for the QIS of an entangled state
of type $(\alpha|00\rangle+\beta|11\rangle)_{aa'}$. The protocol goes as follows: \\
We let Alice possess qubit 1, Bob possess qubit 4 and Charlie 2,3. Alice combines the entangled state $(\alpha|00\rangle + \beta|11\rangle)_{aa'}$ with
her particles and performs a three-partite GHZ measurement.  The outcome of the 
measurement performed by Alice and the entangled state obtained by Bob and Charlie are shown in table \ref{tab3}:

\begin{table}[h]
\caption{\label{tab3} The outcome of the measurement performed by Alice and the state obtained by Bob and Charlie}
\begin{tabular}{|c|c|}
\hline {\bf Outcome of the Measurement } & {\bf State obtained }\\
$\frac{1}{\sqrt{2}}(|000\rangle+|111\rangle)_{aa'1}$&$(\alpha(|000\rangle+|110\rangle)+\beta(|001\rangle-|111\rangle))_{423}$\\
$\frac{1}{\sqrt{2}}(|000\rangle-|111\rangle)_{aa'1}$&$(\alpha(|000\rangle+|110\rangle)-\beta(|001\rangle-|111\rangle))_{423}$\\
$\frac{1}{\sqrt{2}}(|001\rangle+|110\rangle)_{aa'1}$&$(\alpha(|001\rangle-|111\rangle)+\beta(|000\rangle+|110\rangle))_{423}$\\
$\frac{1}{\sqrt{2}}(|001\rangle-|110\rangle)_{aa'1}$&$(\alpha(|001\rangle-|111\rangle)-\beta(|000\rangle+|110\rangle))_{423}$\\
\hline
\end{tabular}
\end{table}

Instead of making a three-particle measurement, Alice can also perform a two particle measurement 
followed by a single particle measurement. Alice can send the outcome of her
measurement to Charlie via two cbits of information. Now, Bob and Charlie can meet up
and convert their state to $(\alpha|000\rangle + \beta |111\rangle)_{423}$ by joint unitary operations 
on their particles. Bob can perform a single partite Hadamard measurement in the basis $\frac{1}{\sqrt{2}}(|0\rangle \pm |1\rangle)_4$ and convey the outcome of his measurements to Charlie via 1 cbit of information. If Bob measures in the basis $\frac{1}{\sqrt{2}}(|0\rangle \pm |1\rangle)$,
then Charlie's state evolves into $(\alpha|00\rangle \pm \beta|11\rangle)_{23}$.
Having, known the outcomes of both their measurements,
Charlie can get the state $(\alpha|00\rangle + \beta|11\rangle)_{23}$ by performing an appropriate unitary operation on his qubits. \\

Let us now consider, the security
of this scheme against an eavesdropper Eve. Assume, that Eve has been able to entangle a qubit to Bob's system.
After Alice performs the measurement, when Bob and Charlie meet up and perform unitary operations on their combined state,
the ancillia is left unchanged. Now, if Bob performs a single partite measurement on his qubit, the Charlie-Eve system collapses into a product state, leaving the unknown qubit information with Charlie. Note that $|C_4\rangle$ cannot be used for the QIS of an arbitrary two qubit state. 

\section{$|C_5\rangle$ for QIS}
The five qubit cluster state is given by,
\begin{equation}
|C_5\rangle = \frac{1}{2}(|00000\rangle + |00111\rangle + |11101\rangle + |11010\rangle).
\end{equation}
Like $|C_4\rangle$, the $|C_5\rangle$ state turns out be an important resource for Quantum communication.
 $|C_5\rangle$ can also be used for teleportation of unknown single and two qubit states. The
superdense coding capacity of this state reaches the Holevo bound, allowing five cbits to be sent using only
three qubits. We shall now proceed to study the usefulness of this state for QIS of single and two qubit systems. 

\subsection{Single qubit state}
We let Alice possess qubit 1, 2, Bob possess qubits 3,4 and Charlie 5. Alice combines the unknown qubit $\alpha|0\rangle + \beta|1\rangle$ with
her qubits and performs a three-particle GHZ measurement and communicates its outcome to Charlie via two cbits. The outcome of the measurement performed by Alice and the entangled state obtained by Bob and Charlie are shown in table \ref{tab6}:

\begin{table}[h]
\caption{\label{tab6} The outcome of the measurement performed by Alice and the state obtained by Bob and Charlie}
\begin{tabular}{|c|c|}
\hline {\bf Outcome of the Measurement } & {\bf State obtained }\\
$\frac{1}{\sqrt{2}}(|000\rangle+|111\rangle)_{a12}$&$(\alpha(|000\rangle+|111\rangle)+\beta(|101\rangle+|010\rangle)_{345}$\\
$\frac{1}{\sqrt{2}}(|000\rangle-|111\rangle)_{a12}$&$(\alpha(|000\rangle+|111\rangle)-\beta(|101\rangle-|010\rangle)_{345}$\\
$\frac{1}{\sqrt{2}}(|011\rangle+|100\rangle)_{a12}$&$(\alpha(|101\rangle+|010\rangle)+\beta(|000\rangle+|111\rangle)_{345}$\\
$\frac{1}{\sqrt{2}}(|011\rangle-|100\rangle)_{a12}$&$(\alpha(|101\rangle+|010\rangle)-\beta(|000\rangle+|111\rangle)_{345}$\\
\hline
\end{tabular}
\end{table}

Now, Bob can perform a two - particle measurement and convey its outcome to Charlie via two cbits of information; thus Charlie's particle evolves into a single-qubit state. For instance, had the Bob-Charlie system evolved into the first state in the table, then the outcome of the measurement performed by Bob and the state obtained by Charlie are shown in table \ref{tab7}:

\begin{table}[h]
\caption{\label{tab7} The outcome of the measurement performed by Bob and the state obtained by Charlie}
\begin{tabular}{|c|c|}
\hline {\bf Outcome of the Measurement } & {\bf State obtained }\\
$\frac{1}{\sqrt{2}}(|00\rangle+|10\rangle)_{34}$&$(\alpha|0\rangle+\beta|1\rangle)_{5}$\\
$\frac{1}{\sqrt{2}}(|00\rangle-|10\rangle)_{34}$&$(\alpha|0\rangle-\beta|1\rangle)_{5}$\\
$\frac{1}{\sqrt{2}}(|01\rangle+|11\rangle)_{34}$&$(\beta|0\rangle+\alpha|1\rangle)_{5}$\\
$\frac{1}{\sqrt{2}}(|01\rangle-|11\rangle)_{34}$&$(\beta|0\rangle-\alpha|1\rangle)_{5}$\\
\hline
\end{tabular}
\end{table}
Having known the outcomes of both their measurements, Charlie can obtain the unknown qubit, by performing a suitable unitary operation on his qubit.\\

We now investigate the security of this protocol. As in the previous cases, let us assume that Eve has managed to entangle
an ancillia to the channel. After, Alice performs a three partite measurement, the combined system of Bob, Charlie
and Eve evolves into an entangled state. Now, if Bob performs a measurement in the Bell basis, then the Charlie-Eve system collapses
into a product state. For instance, had Alice and Bob measured along the first basis, then the Charlie-Eve system collapses
into $(\alpha|0\rangle+\beta|1\rangle)_C(|0\rangle+|1\rangle)_E$. Hence, Charlie gets the unknown qubit information and the protocol is 
secure against these types of eavesdropping attacks. 
\subsection{Arbitrary two qubit state}
In this section, we demonstrate a scheme for QIS of
an arbitrary two qubit state using the five qubit cluster state. 
Alice has an arbitrary two qubit state $|\psi_b\rangle$ which she wants Bob and Charlie to share. 
We now demonstrate the utility of $|C_5\rangle$ for the QIS of an arbitrary
two qubit state.
We let Alice posses particles 1,5, Bob posses particle 2, and 
Charlie posses particles 3 and 4 in $|C_5\rangle$ respectively. Alice first
combines the state $|\psi_b\rangle$ with $|C_5\rangle$ and performs a four - particle von-Neumann measurement and conveys
the outcome of her measurement to Charlie by four cbits of information.
The outcome of the measurement made by Alice and the entangled state obtained by Bob and Charlie are shown in
table \ref{tab9}.

\begin{table}[h]
\caption{\label{tab9} The outcome of the measurement performed by Alice and the state obtained by Bob
and Charlie (The coefficient $\frac{1}{2}$ is removed for convenience.)}
\begin{tabular}{|c|c|}
\hline {\bf Outcome of the Measurement } & {\bf State obtained}\\
\hline
$|0000\rangle+|1001\rangle+|0111\rangle+|1110\rangle$&$\alpha|000\rangle+\mu|011\rangle+\gamma|110\rangle+\beta|101\rangle$\\
$|0000\rangle+|1001\rangle-|0111\rangle-|1110\rangle$&$\alpha|000\rangle+\mu|011\rangle-\gamma|110\rangle-\beta|101\rangle$\\
$|0000\rangle-|1001\rangle+|0111\rangle-|1110\rangle$&$\alpha|000\rangle-\mu|011\rangle+\gamma|110\rangle-\beta|101\rangle$\\
$|0000\rangle-|1001\rangle-|0111\rangle+|1110\rangle$&$\alpha|000\rangle-\mu|011\rangle-\gamma|110\rangle+\beta|101\rangle$\\
$|0001\rangle+|1000\rangle+|0110\rangle+|1111\rangle$&$\alpha|011\rangle+\mu|000\rangle+\gamma|101\rangle+\beta|110\rangle$\\
$|0001\rangle+|1000\rangle-|0110\rangle-|1111\rangle$&$\alpha|011\rangle+\mu|000\rangle-\gamma|101\rangle-\beta|110\rangle$\\
$|0001\rangle-|1000\rangle+|0110\rangle-|1111\rangle$&$\alpha|011\rangle-\mu|000\rangle+\gamma|101\rangle-\beta|110\rangle$\\
$|0001\rangle-|1000\rangle-|0110\rangle+|1111\rangle$&$\alpha|011\rangle-\mu|000\rangle-\gamma|101\rangle+\beta|110\rangle$\\
$|0011\rangle+|1010\rangle+|0100\rangle+|1101\rangle$&$\alpha|110\rangle+\mu|101\rangle+\gamma|000\rangle+\beta|011\rangle$\\
$|0011\rangle+|1010\rangle-|0100\rangle-|1101\rangle$&$\alpha|110\rangle+\mu|101\rangle-\gamma|000\rangle-\beta|011\rangle$\\
$|0011\rangle-|1010\rangle+|0100\rangle-|1101\rangle$&$\alpha|110\rangle-\mu|101\rangle+\gamma|000\rangle-\beta|011\rangle$\\
$|0011\rangle-|1010\rangle-|0100\rangle+|1101\rangle$&$\alpha|110\rangle-\mu|101\rangle-\gamma|000\rangle+\beta|011\rangle$\\
$|0010\rangle+|1011\rangle+|0101\rangle+|1100\rangle$&$\alpha|101\rangle+\mu|110\rangle+\gamma|011\rangle+\beta|000\rangle$\\
$|0010\rangle+|1011\rangle-|0101\rangle-|1100\rangle$&$\alpha|101\rangle+\mu|110\rangle-\gamma|011\rangle-\beta|000\rangle$\\
$|0010\rangle-|1011\rangle+|0101\rangle-|1100\rangle$&$\alpha|101\rangle-\mu|110\rangle+\gamma|011\rangle-\beta|000\rangle$\\
$|0010\rangle-|1011\rangle-|0101\rangle+|1100\rangle$&$\alpha|101\rangle-\mu|110\rangle-\gamma|011\rangle+\beta|000\rangle$\\
\hline
\end{tabular}
\end{table}
It should be noted that each four partite measurement basis in table V can be further broken down into Bell and single partite measurements,
making the scheme experimentally feasible. For instance, the first measurement basis can be written as,
\begin{equation}
(|\psi_+\rangle (|0\rangle+|1\rangle) + |\psi_{-}\rangle (|0\rangle-|1\rangle)) |0\rangle + (|\phi_{-}\rangle (|0\rangle-|1\rangle) +
|\phi_+\rangle (|0\rangle+|1\rangle)) |1\rangle,
\end{equation}
where $|\psi_+\rangle$, $|\psi_{-}\rangle$, $|\phi_+\rangle$ and $|\phi_{-}\rangle$ refers to the Bell states.
Bob performs a measurement in the basis $\frac{1}{\sqrt{2}}(|0\rangle\pm|1\rangle)$ and communicates the outcome of his measurement to Charlie who then, performs local unitary transformations to get the state $|\psi_b\rangle$. 
For instance, had the Bob-Charlie system been the first state, then, after Bob's measurement, Charlie's state collapses to $(\alpha|00\rangle
\pm\mu|11\rangle\pm\gamma|10\rangle+\beta|01\rangle)$, which can be converted to $|\psi\rangle$, by an appropriate Unitary
operation. This completes the protocol for QIS of an arbitrary two qubit state using $|C_5\rangle$. The security of this protocol
against eavesdropping attacks will require further investigation. Nevertheless, the protocol is significant, as the threshold number
of qubits that an entangled channel should possess for the QIS of an arbitrary two qubit state in the case, where
Bob and Charlie need not meet up, is five.
\section{Conclusion}
Cluster states are one of the most widely discussed multi-partite states 
in 	Quantum information theory. Applications such as one-way quantum computing, teleportation and superdense coding,
using the cluster states have already been discussed in detail. In this paper, we show the efficacy of cluster states for Quantum information splitting. We discussed different scenarios in which four and five partite cluster states can be used
for the QIS of single and two qubit states. The schemes considered are secure against certain types of eavesdropping attacks. Any experimental ventures
for splitting up an arbitrary two qubit information can make use of our scheme. 
A detailed crypto-analysis of our work and an investigation of  robustness of all the protocols
considered in this paper against particle loss is currently under investigation. We are also looking forward to generalize our schemes for QIS of an arbitrary two qubit state using the $N$ dimensional cluster state as an entangled channel. Further, it remains to be proved that, one can devise $(N-4)$ protocols for this purpose.


\begin{thebibliography}{26}
\bibitem{Bennett} C. H. Bennett, G. Brassard, C. Crepeau, R. Jozsa, A. Peres, and W. K. Wootters,
 \PRL{70}{1895}{1993}.
\bibitem{tele1} D. Bouwmeester, J. W. Pan, K. Mattle, M. Eibl, H. Weinfurter, and A. Zeilinger
\NAT{390}{575}{1997}.
\bibitem{Bsm} Q. Zhang, A. Goebel, C. Wagenknecht, Y. A. Chen, B. Zhao, T. Yang, A. Mair, J. Schmiedmayer and J. W. Pan,
\NAT{2}{678}{2006}.
\bibitem{Chuang28} L.D. Chuang, and C.Z. Liang, \CTP{47}{464}{2007}.
\bibitem{Rigolin28} G. Rigolin, \PRA{71}{032303}{2005}.
\bibitem{Gorbo} V. N. Gorbachev and A. I. Trubilko, \JETP{91}{894}{2000}.
\bibitem{Shi} B. S Shi, Y. K. Jiang and G. C. Guo, \PLA{268}{161}{2000}.
\bibitem{Deng28} F.G. Deng, C. Y. Li, Y. S. Li, H. Y. Zhou and Y. Wang, \PRA{72}{022338}{2005}.
\bibitem{Sreramanpr} S. Muralidharan and P. K. Panigrahi, eprint quant-ph/arXiv:0802.3484v1 {2008}.
\bibitem{Gotit} D. Gottesman, \PRA{61}{042311}{2000}.
\bibitem{Hillery} M. Hillery, V. Buzek, and A. Berthiaume, \PRA{59}{1829}{1999}.
\bibitem{Band} S. Bandyopadhyay, \PRA{62}{012308}{2000}.
\bibitem{Tittel28} W. Tittel, H. Zbinden, and N. Gisin, \PRA{63}{042301}{2001}.
\bibitem{Christ} C. Schmid, P. Trojek, M. Bourennane, C. Kurtsiefer, M. Zÿukowski, and H. Weinfurter, \PRL{95}{230505}{2005}.
\bibitem{Gaer}  S. Gaertner, C. Kurtsiefer, M. Bourennane, and H. Weinfurter, \PRL{98}{020503}{2007}.
\bibitem{Zheng}S. B. Zheng, \PRA{74}{054303}{2006}.
\bibitem{Wqed} W. H. Zhi, Y. Z. Biao, S. W. Jun, Z. Z. Rong and H. J. Min, \CTP{49}{1165}{2008}.
\bibitem{Wqed2} Y. X. Mei, G. Y. Jian, M. L. Zhen and Z. B. An, \Chtp{17}{462}{2008}.
\bibitem{Karl} A. Karlsson, M. Koashi, and N. Imoto, \PRA{59}{162}{1999}.
\bibitem{Cleve} R. Cleve, D. Gottesman, and H. K. Lo, \PRL{83}{648}{1999}.
\bibitem{quditqis}  W.J. Kim, S.H. Cha, S.W. Lee, and J. Lee \JKP {48}{1218}{2006}.
\bibitem{Deng} F. G. Deng, X. H. Li, C. Y. Li, P. Zhou, and H. Y. Zhou, \PRA{72}{044301}{2005}.
\bibitem{Sreraman} S. Muralidharan and P. K. Panigrahi, \PRA {77}{032321} {2008}.
\bibitem{Robert} R. Raussendorf and H. J. Briegel, \PRL{86}{5188}{2001}.
\bibitem{Hein} M. Hein, W. Dur, and H. J. Briegel, \PRA{71}{032350}{2005}.
\bibitem{Walther} P. Walther, K. J. Resch, T. Rudolph, E. Schenck, H. Weinfurter, V. Vedral, M. Aspelmeyer  and A. Zeilinger, \NAT{86} {434}{169} {2005}.
\bibitem{Bai} Y. K. Bai and Z. D. Wang, eprint quant-ph/0709.4642v1.
\bibitem{MEMS} D. Liu, X. Zhao, and G. L. Long, eprint quant-ph/0705.3904v4.
\bibitem{Pankaj} P. Agrawal and B. Pradhan, eprint quant-ph/0707.4295v2.
\bibitem{Lu} C. Y. Lu, X. Q. Zhou, O. Gühne, W. B. Gao, J. Zhang, Z. S. Yuan, A. Goebel, T. Yang, J. W. Pan,
\NAT{3}{91}{2007}.
\bibitem{Hans} H. J. Briegel and R. Raussendorf, \PRL{86}{910}{2001}.
\bibitem{Sch} D. Schlingemann and R. F. Werner, \PRA{65}{012308}{2001}.
\bibitem{4pati} B. Pradhan, P. Agrawal, and A. K. Pati, eprint quant-ph/0705.1917v1.
\bibitem{mono} V. Coffman, J. Kundu and W.K. Wootters \PRA{61}{052306}{2000}.




\end{thebibliography}
\end{document}